# Rare-earth monosulfides
# as durable and efficient cold cathodes*

Marc Cahay[a)]

School of Electrical and Computing Systems, University of Cincinnati, Cincinnati, OH 45221

Steve B. Fairchild, Larry Grazulis

Air Force Research Laboratory, Materials and Manufacturing Directorate, AFRL/RXPS, Wright-Patterson AFB, OH 45433

Paul T. Murray

Research Institute, University of Dayton, Dayton, OH 45469

Tyson C. Back

Universal Technology Corporation, 1270 N. Fairfield Rd, Dayton, OH 45432

Punit Boolchand

School of Electrical and Computing Systems, University of Cincinnati, Cincinnati, OH 45221

Vincent Semet, Vu Thien Binh

Equipe Emission Electronique, LPMCN, University of Lyon 1, Villeurbanne 69622, France

Xiaohua. Wu, Daniel Poitras, David J. Lockwood

Institute for Microstructural Sciences, National Research Council, Ottawa, Ontario K1A OR6, Canada

Fei Yu, Vikram Kuppa

School of Energy, Environmental, Biological, and Medical Engineering, University of Cincinnati, Cincinnati, OH 45221

In their rocksalt structure, rare-earth monosulfides offer a more stable alternative to alkali

metals to attain low or negative electron affinity when deposited on various III-V and II-

VI semiconductor surfaces. In this article, we first describe the successful deposition of Lanthanum Monosulfide (LaS) via pulsed laser deposition on Si and MgO substrates and alumina templates. These thin films have been characterized by X-ray diffraction, atomic force microscopy, high resolution transmission electron microscopy, ellipsometry, Raman spectroscopy, ultraviolet photoelectron spectroscopy and Kelvin probe measurements. For both LaS/Si and LaS/MgO thin films, the effective work function of the submicron thick thin films was determined to be about 1 eV from field emission measurements using the Scanning Anode Field Emission Microscopy technique. The physical reasons for these highly desirable low work function properties were explained using a patchwork field emission model of the emitting surface. In this model, nanocrystals of low work function materials having a <100> orientation perpendicular to the surface and outcropping it are surrounded by a matrix of amorphous materials with higher work function. To date, LaS thin films have been used successfully as cold cathode emitters with measured emitted current densities as high as 50 A/cm$^2$. Finally, we describe the successful growth of LaS thin films on InP substrates and, more recently, the production of LaS nanoballs and nanoclusters using Pulsed Laser Ablation.

## I. INTRODUCTION

Typically, low work function surfaces are generally highly chemically reactive. As a result, cold cathodes made of these materials are very unstable. Beginning in 2001, we have studied rare-earth (RE) monosulfide bulk samples and thin films for use in the development of a new class of field emitters that turned out to be very reliable and durable[1-4]. RE compounds do not suffer from all the limitations of cesiated surfaces. A summary of some of the material properties of RE monosulfides in their rocksalt cubic form is given in Table I. When extrapolated from high-temperature measurements[5], the work function (WF) of these compounds is predicted to be quite small at room temperature (around 1 eV). It is therefore expected that these materials can be used to reach negative electron affinity (NEA) when deposited on p-type doped semiconductors. For instance, LaS has a lattice constant (5.854 Å) very close to the lattice constant of InP (5.8688 Å) and NdS has a lattice constant (5.69 Å) very close to the lattice constant of GaAs (5.6533 Å). Since the room temperature WF of LaS (1.14 eV) and NdS (1.36 eV) are respectively below the band gap of InP (1.35 eV) and GaAs (1.41 eV), NEA can therefore be reached at InP/LaS and GaAs/NdS interfaces using heavily p-type doped semiconductors. Recently, we confirmed this result by means of a first-principle electronic-structure method based on a local-density approximation to density-functional theory[6]. This analysis predicted a 0.9 eV and 1.1 eV WF for LaS[7] and NdS[8], respectively, at low temperature. Two other important features of the face-centered-cubic form of the rare-earth compounds listed in Table I are the fairly large melting temperature (about 2000°C) and their fairly good electrical resistivity (a few tens of μΩ-cm).

Our work was motivated by the original proposal by Walter Friz for an InP/CdS/LaS cold cathode emitter concept, which we refer to hereafter as the *Friz cathode*[9]. The architecture of the structure is shown in Fig. 1. A schematic energy band diagram along the growth direction throughout the proposed structure is shown in Fig. 2. The main elements in the design and functioning of such an emitter are: (1) a wide bandgap semiconductor slab equipped on one side with a heavily doped n-type InP substrate that supplies electrons at a sufficient rate into the conduction band and (2) on the opposite side, a thin semimetallic LaS film that facilitates the coherent transport (tunneling) of electrons from the semiconductor conduction band into vacuum. As shown in Fig.1, an array of Au contacts is defined on the surface of the LaS thin film to bias the structure. The bias is applied between the InP substrate and the metal grid with emission occuring from the exposed LaS surface.

In the Friz cathode, the choice of a LaS semimetallic thin film grown on nominally undoped CdS is quite appropriate since the lattice constant of CdS (5.83 Å) is very close to the lattice constant of LaS (5.85 Å) in its cubic crystalline form. Additionally, LaS is expected to have a quite low temperature work function (1.14 eV)[5], a feature when combined with the large energy gap (2.5 eV) of CdS leads to NEA of the semiconductor material. In fact, we have studied the electronic structure of the CdS/LaS interface by means of a first-principles electronic-structure method based on a local-density approximation to density-functional theory[7]. The extrapolated 1.14 eV low work function of LaS was reproduced by that theory. It was found that NaCl structured layers of LaS should grow in an epitaxial way on a CdS substrate with a ZnS structure[7].

In 1996, Mumford and Cahay modeled the Friz cathode and showed that with a small forward bias (< 20 V), some of the electrons originating from the InP substrate and tunneling through the CdS layer are captured in the LaS thin film eventually leading to an effective reduction in the semimetallic thin film[10]. This leads to a substantial increase in the Friz cathode emission current. Current densities of several 100 A/cm$^2$ could be achieved with this cathode. With enough forward bias on the emitter fingers, electrons emitted from the InP substrate tunnel through the CdS layer and are emitted into vacuum as a result of NEA at the CdS/LaS interface. However, there can be partial trapping of electrons by the LaS semimetallic thin film, which can lead to a lateral current flow and current crowding[11,12] and self-heating effects[13] in the cathode.

This paper is organized as follows. We first review our preliminary work on making LaS targets and the growth of LaS thin films on Si and MgO substrates using Pulsed Laser Deposition (PLD). Various characterization tools were used to study the LaS thin films including: X-ray diffraction (XRD), atomic force microscopy (AFM), high resolution transmission electron microscopy (HRTEM), Raman spectroscopy, and ellipsometry. The thin film WF was determined in situ by Ultraviolet Photoelectron Spectroscopy (UPS) and compared with the results of Kelvin Probe (KP) measurements in air. These WF values was compared to the results for the effective WF values extracted from field emission (FE) of the thin films recorded using scanning anode field emission microscopy (SAFEM). The low effective WF (~ 1 eV) of the films was interpreted in terms of a recently developed patchwork FE model of the emitting surface. The main ingredients of the patchwork FE model and its practical implications are emphasized. Finally, we describe the successful growth of LaS thin films on InP substrates and, more recently, the production of LaS nanoballs and nanoclusters using Pulsed Laser Ablation (PLA).

## II. EXPERIMENTAL TECHNIQUES

In our PLD experiments[1-4] the target to substrate distance was set equal to about 10 cm and the vacuum chamber base pressure was typically a few $10^{-8}$ Torr. A pulsed Lambda Physik LPX 305 excimer laser operating at a wavelength of 248 nm and a repetition rate up to 10 Hz was used for the deposition. The beam spot size on target was about 4 mm by 8 mm and the laser energy was estimated to be between 300 and 400 mJ/pulse. During deposition the target was rotated on axis while galvanometers were used to raster the laser beam in a uniformly random pattern over the surface of the target. This configuration produces a uniform laser plume by preventing uneven erosion of the target.

XRD $2\Theta$ scans were performed using a PanAlytical X'Pert Pro x-ray diffraction system in which the incident angle was set equal to 5°. Surface images were obtained by AFM using a Pacific Nanotechnology Nano-R SPM (model no. 0-200-002) operating in vibrating mode. Line scans were performed to measure the surface roughness and the grain size of the films was extracted using the Scherrer formula[14].

The LaS films were characterized with Staib multi-technique surface analysis system equipped with a DESA 150 double-pass cylindrical mirror analyzer. The background pressure in the system was 3 x $10^{-10}$ Torr. The binding energy scale was calibrated against the Fermi edge of a sputter cleaned Au sample. UPS data were acquired using the Specs 10/35 He I (21.22 eV) gas discharge lamp. Resolution was determined from the width of Au Fermi edge, which came out to be 250 meV. The sample was negatively biased in order to determine the true work function of the sample. The WF of the thin films was also measured in situ using a McCallister KP-6500 Kelvin Probe with a stainless steel tip with a 5 mm diameter. The procedure involved first determining the WF of the tip using UPS. The WF tip value was found to be 4.8 eV. The absolute value of the film WF was then determined by the difference between the KP tip and the sample as recorded in the KP experiments.

A variable-angle spectroscopic ellipsometer (VASE, J.A. Woollam Co. Lincoln NE) was used for the characterization of the optical constants of the film in the near UV and visible wavelength range.

To investigate the crystalline texture of the films, a <011> cross-section TEM sample was prepared following standard procedures. Two bars were cut out of a wafer, and were glued together with the film sides face-to-face to make the central part of a 3 mm diameter cross-section disc. Then the disc was mechanically thinned to ~ 100 μm thick. The thinned disc was dimpled from both sides with 3 μm diamond paste until the center of the disc was ~ 20 μm thick, and then polished from both sides with 1 μm diamond paste to get a very smooth surface. The final thinning until perforation was conducted using Ar ion-milling from both sides using an ion beam angle of 8°, and a gun voltage of 6 kV. The TEM sample was examined in a Philips EM430T operating at 250kV.

The SAFEM technique was used to measure the current-voltage (I-V) characteristics at different surface locations and at different temperatures of a LaS thin film with an area of about a 1 cm$^2$. For one location, the full set of measured I-V characteristics (total measured current versus applied voltage) for different values of Z, the distance between the cathode surface and the probe ball, were then analyzed in order to extract the current density J versus actual applied field F (J-F) data[15,16].

Raman measurements were performed at 295 K employing 1 mW of 488 nm light from a Picarro diode-pumped solid state laser for excitation in a backscattering geometry using the confocal microscope with a 100× objective on a Jobin-Yvon T64000 triple spectrometer operated in the subtractive mode. The incident laser light was focused on the LaS layer.

## III. PRIOR EXPERIMENTAL WORK

### A. *RE target preparation*

In 2001, Cahay and coworkers reported the growth of a bulk sample of LaS using a two-step sesquisulfide route[17]. To grow the cubic phase of LaS, we first reacted equimolar portions of the rare-earth sesquisulfide La$_2$S$_3$ and rare-earth element La at 1800 $^o$C for two hours in a high-temperature furnace from Thermal Technologies, Inc. A carbon reduction process during the high temperature annealing portion of the growth process was shown to greatly reduce the presence of oxysulfide impurity phases (La$_2$O$_2$S and Nd$_2$O$_2$S) in the samples. A picture of a LaS pellet prepared in a high-temperature induction furnace from Thermal Technologies, Inc., is shown in Fig. 3(a). Figure 3(b) shows an optical micrograph of a 30 µm² area inside a pellet showing a staircase-like growth pattern. Several of these pellets were prepared using the carbon reduction technique until enough material was produced to form a 2-inch diameter 1/8-inch thick target, which was then sintered in the high-temperature furnace. The resulting target was then used for PLD of LaS thin films. A comparison of XRD spectra taken on a bulk LaS sample before and after the carbon reduction process is shown in Fig.4.

## *B. PLD experiments*

### *1. LaS/Si thin films*

In the past, we reported the successful growth by PLD of nano-crystalline LaS thin films on Si (100) substrates at temperature ranging from room temperature to 100°C[1,2]. As shown in Fig.5, these films are golden yellow in appearance with a mirror-like surface morphology and a sheet resistance of around 0.1 Ω/square, as measured using a 4-probe measurement technique. For a film grown at a rep rate of 40 Hz with a substrate temperature of 100°C and no background gas pressure, the growth rate was estimated to be about 20 nm/min. An AFM scan of this film over a 1 µm² area of a 1 micron thick film is shown in Fig. 6. The root-mean-square variation of the surface roughness over the 1 µm² area was found to be 1.74 nm. Figure 7 shows a XRD scan of this film revealing the successful growth of the cubic rocksalt structure with a lattice constant of 5.863(7) Å, which is close to the bulk LaS bulk value of a = 5.857(2) Å. The observation of many Bragg reflection peaks suggests that the film is polycrystalline, which is to be expected because of the large lattice mismatch (about 8%) between the lattice constant of the LaS rocksalt phase and of the Si substrate (a=5.41 Å). In addition, the Bragg peaks are very broad suggesting the presence of polycrystalline grains of different sizes and potentially the presence of amorphous regions in the film. Using Scherrer's equation, the size of the nanocrystalline grains was found to be about 13 nm. This value agrees well with cross-section TEM analysis of the LaS films, as shown in Fig.8. Extensive TEM analysis reveals the amorphous and crystalline regions comprised approximately 60% and 40% of the film, respectively[1,2].

## 2. *LaS/MgO thin films*

LaS thin films were grown by PLD on (001) MgO substrates at an elevated substrate temperature and in a background gas of $H_2S$. The goal of this research was to prepare crystalline, epitaxial films of LaS with (100) orientation capable of delivering higher emission current density than reported for the largely amorphous LaS/Si thin films. The approach is based on PLD at high temperature in a background gas of $H_2S$ following the work of Piqué[18,19]. Single crystal MgO (001) with the NaCl structure was selected as the substrate of choice: MgO has a lattice constant of 4.213 Å. Cubic LaS has a lattice constant of 5.857 Å leading to a lattice mismatch of 1.69 % when aligned with the unit cell diagonal of MgO (5.957 Å). It was found that a substrate temperature of at least 400 °C coupled with a low laser repetition rate and at least 20 mTorr of $H_2S$ were needed to produce crystalline, cubic LaS as evidenced by XRD[3]. As shown in Fig. 9(a), for a LaS/MgO thin film grown at 400 °C in 24 mTorr of $H_2S$ with a laser repetition rate of 4 Hz for 30 min, the film surface is composed of grain-like features with an average size of approximately 34 nm and the root-mean-square variation of the film surface roughness measured over a 2x2 $\mu m^2$ area by AFM was found to be about 1.5 nm. XRD data indicate that the average size of the nanocrystalline grains in the film is about 26 nm, which is about twice the size of the grains found in LaS thin films deposited at room temperature on Si.

For comparison, Fig. 9(b) shows an AFM image of a LaS/Si sample over a 2x2 $\mu m^2$ area grown using the same PLD parameters and no $H_2S$ background gas with the substrate held at room temperature. Figure 9(b) indicates that this film is much smoother with a RMS variation in the surface roughness equal to 1.13 Å. The AFM scans demonstrate an abundance of grain-like features in the LaS/MgO thin films leading to enhanced field emission properties for these films. As discussed below, the field emission (FE) properties of the LaS/MgO films have been characterized by the SAFEM technique and interpreted in terms of a patchwork FE model. The FE data indicate that there is roughly a 7 times increase in emission area due to (100) patch areas outcropping the surface for LaS/MgO films compared to LaS/Si thin films.

More recently, we have carried a thorough investigation of the growth of LaS thin films by PLD on (001) magnesium oxide (MgO) substrates in a background gas of $H_2S$ for the purpose of optimizing their texture and crystallinity[20]. A large variety of films were grown while varying the laser repetition rate (from 3 to10 Hz), the temperature of the substrate (from 300 to 500 °C), and the partial pressure of $H_2S$ (from 25 to 50 mTorr).

Films grown at 500 °C with a H$_2$S background pressure of 45 mTorr at a laser repetition rate of 8 Hz produced the LaS films with the largest grains size, averaging 293 nm. The XRD pattern of these films revealed that their orientation is predominantly (100). As shown Fig.10, images of the surface of these films showed large, flat, plate-like platelets. This is a dramatic change in grain structure compared to LaS films grown at a 400°C and at a lower H$_2$S pressure of 24 mTorr, as shown in Fig. 9(a).

### C.  *WF Measurements*

For UPS analysis, the LaS films were Ar etched at 1.5keV for 20s before data acquisition.  The WF of the samples was determined by substracting the width of the spectrum from the energy of the He I line (21.22 eV). The width of the spectrum was taken from zero binding energy (Fermi edge) to the onset of the secondary edge[21]. Figure 11 shows the UPS spectra of the LaS film. The sharp feature at low kinetic energy is the secondary edge cut-off. From this feature a WF of 2.9 eV was determined. The inset shows a blow-up region around the Fermi energy. This value agrees well with the best work function of 2.6 eV recorded in air using a KP on a LaS pellet produced by an arc melting process[22]. For LaS/Si and LaS/MgO thin films, the WF values determined in air using a KP technique ranged from 3.7-4.7 eV[23].  These higher values are believed to be due to partial oxidation of the top surface of the film after exposure to air. Next, we compare these WF values to those extracted from FE measurements using the SAFEM technique.

### D. SAFEM Measurements

The main reason for of surface instabilities and breakdown of low work function cathodes (such as cesiated surfaces) results from a non controlled surface diffusion at the cathode surface induced byl heating due to the field emission current. This thermal run-away process is amplified by the presence of adsorbates on the cathode surface. RE monosulfide thin films, such as LaS/Si and LaS/MgO, in their rocksalt structure, offer a stable alternative as low work cathode material because of the patchwork nature of their field emitting surface, as will be discussed next.

To measure the field emission characteristic from LaS planar cathodes, we have scanned an anode probe located at micrometer distance over the cathode surface. Such an analytical device, called Scanning Anode Field Emission Microscope (SAFEM), has been fully described earlier[24]. The corresponding analytical procedure to extract, from the measured I-V data, the specific parameters of the cathode and in particular its surface barrier or WF have also been discussed previously[3,4,25]. The measurements have been done with different LaS planar cathodes, and the results were given in detail in recent articles where we studied field emission from LaS layers deposited on Si substrates[1,2,25], LaS thin films grown on MgO substrates[3], and LaS thin films deposited on self-assembled nanostructures[4]. Hereafter, we summarize the main findings:

(1) Stable and reversible field emission characteristics were obtained from a threshold of the applied electric field of ~230 V/µm and for current densities less than 0.3 and 7 A/cm², respectively for Si and MgO substrates.

(2) The measured surface barrier extracted from Fowler-Nordheim analysis of the I-V characteristics was in the range of 0.7 to 1 eV.

(3) The comparison between a LaS flat film and layers with self-assembled nanodomes indicated a straightforward influence of the field enhancement factor γ of the nanoprotrusion (Fig. 12). Considering the geometry of the nanodomes, the value of 3.6 for the field enhancement is in good agreement with the well known value of 3 for a hemisphere on a plane[26].

(4) The comparison between LaS thin films deposited on Si and on MgO showed two main features. First, the surface barrier in both cases was the same and was in the order of 1 eV. Second, the field emission current for LaS on MgO was 7 times greater than for LaS on Si (Fig. 13). This specificity has to be placed in perspective with the areal density of the (100) oriented nanopatches which was found to be 7 time larger at the surface of the LaS/MgO thin films compared to those for the LaS/Si thin films. A HRTEM picture of the surface of a LaS/Si is shown in Fig. 14(top) where nanopatches surrounded by amorphous material can clearly be seen. A schematic of the (100) oriented nanopatches outcropping the surface is shown in Fig. 14(bottom).

(5) There was a limiting current beyond which the low surface barrier (1 eV) was destroyed. Beyond this threshold, further FE measurements gave an extracted value of 2.8 eV from the Fowler-Nordheim plot for the WF in agreement with Kelvin probe measurements[25].

# IV. PATCHWORK FIELD EMISSION MODEL

The analysis of the experimental results listed above must explain the following characteristics:

(1) Why is there a difference in the WF values recorded with a Kelvin probe and those extracted from the SAFEM technique? 2.8-3 eV compared to 1 eV, respectively.

(2) The effective WF of 1 eV extracted from the SAFEM measurements was obtained without the need for extensive surface cleaning treatment, except for a gentle acetone rinse. So why did we observe a surface with such a low barrier with apparently very little reaction with molecular species in the surrounding gases?

(3) What is the cause for the increase in the WF from 1 eV to 2.8 eV when the FE current goes through a threshold value?

Over the last few years, we have developed a patchwork FE model of LaS thin films[27]. Our analysis was based on the following two assumptions, resulting from the nanocrystalline structure of the LaS films described above:

(1) The LaS nanocrystals presented low WF facets, in the order of 1 eV. This value is consistent with the (100) crystallographic orientation of LaS[6].

(2) Some of these (100) facets were outcropping the surface of the LaS thin films. These low WF nanocrystalline patches are surrounded by areas with an average WF of 2.8 eV.

The idea of patchwork field emission was first formulated by Herring and Nichols in 1949 to interpret the field emission from micron size areas in thin film cathodes[28]. We have extended this model to the nanoscale region to explain the FE characteristics of LaS thin films. In Figure 15(a), we show a schematic of a circular area representing the surface of a <100> oriented facet of a nanocrystal of LaS at the top of a thin film grown on a Si or MgO substrate, as shown in Fig. 14(a).

In our model, the <100> oriented nanocrystal is shown as a circular patch of radius R with a low WF ($\Phi_1 \sim 1$ eV) and surrounded by an amorphous phase with a larger WF $\Phi_2$ (2.8-3 eV). The latter value was determined using a Kelvin probe measurement of the WF. It is expected that the interface between the low and high WF regions is not abrupt, but extends over a contact zone of width $L_C$, as shown in Fig. 15(a). At equilibrium, the Fermi level $E_f$ in the patch and surrounding area must be aligned (see Fig. 15(b)). Therefore, the potential energy profile along a line going from the center of the patch towards the outside in the plane of the LaS film must look as shown in Fig. 3(b). An estimate of the electric field $F_{patch}$ over the patch is therefore given by[24]

$$e\, F_{patch} = (\Phi_2 - \Phi_1)/R, \qquad (1)$$

which is of the order of $10^7$ V/cm.

With an anode placed on top of the thin film, the potential energy profile in the vacuum region was calculated by solving the Laplace equation within the zero emitted current approximation, i.e., assuming that no electron beam is injected from the patches with low WF. The energy potential contours are shown in Fig. 16 for different values of the applied electric field $F_{app}$ across the vacuum gap[25].

For low values of the applied electric field, the presence of the patch is hindered within a distance of a few nm from the surface by the electrostatic potential generated by the large WF surrounding area (Fig.16(a)). Practically, this suggests that WF measurements by field free macroscopic techniques, (i.e., Kelvin probe, UPS, thermionic and photoemission measurements), operating with probe distances more than 100s of nm away from the surface, cannot detect the presence of low work-function patches if they are of nanometer sizes, typically of the order of 10 nm or less. This explains the discrepancy between our WF measurements by Kelvin probe and FE characteristics. Moreover, the potential screening of the low WF nanopatches by the surrounding high WF medium was sufficiently efficient to prevent strong surface chemical adsorption occurring on as-grown pristine LaS cathodes, therefore allowing the possibility to obtain FE without any prior thermal or sputtering treatments.

Eventually, the electrostatic opening of the patch occurs when the contribution of the externally applied field counterbalances the patch field estimated by Eq. (1), as shown in Fig.16. Therefore, the potential distribution over the patch is dependent on the patch geometry, as it is illustrated in Fig. 17. This means that FE from the patch cathode will be facilitated with an increase in the dimension of the patch, reaching asymptotically the classical behavior of two independent areas having, respectively, WFs of 1 eV and 2.8 eV. However, an increase in the patch dimension in order to increase the FE current must be balanced by the electrostatic potential protection of the patch by the high WF surrounding area in order to keep the patch surface free from adsorbates. This limits the size of the patch to an upper value of one hundred nm.

From the progressive opening of the patch, the I-V characteristics from the patchwork cathode will first be governed by the progressive field emission from the patch until the beginning of the field emission from the surrounding region. Considering the difference in WF values between these two zones, the 1 eV patch region is largely in the ballistic regime before noticeable current can be extracted from the surrounding 2.8 eV WF surface. This is illustrated in Fig. 18, in which we have plotted simultaneously the I-V characteristics of the patchwork cathode, and of uniform surface cathodes of 1 eV and 2.8 eV WF.

In Fig.18, the saturation zone of the 1 eV cathode corresponds to the ballistic regime for field emission, i.e., the top of the surface barrier is below the Fermi level. One can notice that before that ballistic threshold, the variation of the patch cathode is very comparable to the 1 eV cathode thus proving that the emission current comes mainly from the patch area. Theoretically, when the 1 eV area reaches the ballistic regime the variation of the field emission current is then governed mainly by the FE current variation of the surrounding 2.8 eV area. This is illustrated in Fig. 18, where for $1/F < 10$ Å/V, the I-V plots for the patchwork cathode and the uniform 2.8 eV cathode have identical slopes. In an actual experiment, once the ballistic regime of operation is reached, the FE current density extracted from the patches can become so important that the patches can be destroyed by local field emission induced thermal effect, resulting in a uniform 2.8 eV cathode.

## V. RECENT EXPERIMENTAL WORK

### A. *Growth of LaS thin films on InP substrates*

LaS thin films were grown by PLD on (100) oriented InP substrates at room temperature using the following PLD parameters: rep rate = 4 Hz, laser energy = 850 mJ/pulse, duration = 30 min. Figure 19 shows the X-ray diffraction pattern recorded at three different grazing angles of incidence. Miller indices (hkl) of the reflections from the cubic rocksalt phase of LaS are identified. The Bragg reflections peaks are narrower than those for a LaS thin film on Si substrate, as shown in Fig.7, indicating a much higher level of crystallinity in the Las/InP film due to the closer lattice-match between the two materials.

Figure 20 shows there are two LaS layers on the InP substrate, and the thicknesses of LaS(1) and LaS(2) are 52 nm and 121 nm respectively. The surface and interface between LaS(1) and LaS(2) are flat in some areas (Fig.20(a)), but undulated in other areas (Fig.20(b)). A (111) planar lattice defect indicated by a white arrow in Fig. 20(a) was observed in LaS (2). Selected area diffraction patterns from the LaS film (Fig. 20(c)) reveal that the LaS film has a poly-crystalline nature.

Ellipsometric spectra $\Psi(E)$ and $\Delta(E)$ measured from LaS on InP are shown in Figure 21(a) and (b), respectively. Given the small size of the probed sample, the area of the incident beam used in ellipsometry had to be reduced, which may account for the large noise that is seen in the data, particularly in the UV part of the spectra where the signal was particularly weak. To extract the complex permittivity dispersion curves $\varepsilon_1(E)$ and $\varepsilon_2(E)$ [shown in Figure 21(c)], we attempted to fit the data using information from the TEM cross-section of the LaS film on InP (see Fig.20); the model used for fitting is described in Table 2, with a rough surface layer partially oxidized, and two different layers of LaS on InP. The dispersion curve of both LaS layers were based on a Drude-Lorentz type of oscillators:

$$\varepsilon = \varepsilon_\infty - \frac{A_D \Gamma_D}{E^2 + i\Gamma_D E} + \frac{A_L \Gamma_L}{E_L^2 - E^2 - i\Gamma_L E} = \varepsilon_1 - i\varepsilon_2.$$ (Eq. 2)

We found that (i) a rough and oxidized surface layer of 35 nm was needed for the fitting (which is in good agreement with the TEM observations), while (ii) parameters from a bottom LaS layer had little influence on the fitting. The permittivity found for the 134-nm thick LaS layer is shown in Figure 21(c); the incident light energy value for which $\varepsilon_1$ equals zero (2.53 eV) corresponds to the plasma frequency below which the material starts behaving as a metal. Figure 21(d) shows the estimated penetration depth of the light into the LaS film based on the permittivity dispersion; it can be seen that below 3.5 eV, the light is not expected to reach the bottom LaS layer found on the TEM micrograph, which explains the little impact of this layer in our model, and is a typical metallic behavior.

Using a Drude model only (first two terms of Eq. (2)), we re-fitted the portion of the data corresponding to the metallic region (energy values below the plasmon frequency), and estimated a carrier concentration and a mobility of $8.44 \times 10^{22}$ cm$^{-3}$ and 1.42 cm$^2 \cdot$V$^{-1} \cdot$s$^{-1}$, respectively (assuming a effective mass m*=1.3) for the 134-nm thick LaS layer., and a resistivity of $5.22 \times 10^{-5}$ Ω·cm$^{-1}$ for that same layer.

The Raman spectra of LaS/InP films exhibit weak broad features extending out to 600 cm$^{-1}$, as illustrated in Fig. 22. Figure 21(d) shows that the Raman excitation light at 488 nm has a penetration depth of only 30 nm in the film. The results obtained are generally similar, but not in detail, to those found from Raman measurements on bulk pure LaS samples (see Fig.22), confirming that the LaS films are of the pure cubic form. Strong peaks are observed at about 86 and 108 cm$^{-1}$, with weaker features at 178, 198, 261, 301, 347, 448, and 518 cm$^{-1}$.

One triply-degenerate optical phonon mode, split into longitudinal (LO) and transverse (TO) components, is expected for rare-earth monochalcogenides possessing the rocksalt crystal structure. First-order Raman scattering from phonons in such crystals is symmetry forbidden and usually a weak second-order Raman spectra is observed reflecting the two-phonon density of states (DOS). However, previous Raman studies of other rare earth chalcogenides[29,30] have shown that disorder introduced by defects such as cation and/or anion vacancies[29] can induce a first-order Raman spectrum reflecting the one-phonon DOS. The TEM analysis of our samples shows that structural disorder due to the formation of nanocrystals separated by amorphous material is also present. Thus the Raman spectrum of our LaS films can be expected to show both disorder-induced first-order as well as second-order Raman features. In fact, our results for LaS films on InP resemble those obtained previously for LaS films on Si[2]. However, a closer examination of Fig.22 shows that the Raman spectra of the present samples are quantitatively different from that of bulk LaS[31]: the Raman peaks above ~150 cm$^{-1}$ are much weaker and more diffuse than those found in bulk LaS and no sharp features are seen at 106, 178, and 189 cm$^{-1}$. This is consistent with disorder in the LaS films.

The stronger features at 86 and 108 cm$^{-1}$ are clearly due to disorder induced first-order Raman scattering from acoustic (A) phonons[2,29,30,31] while the peak at 261 cm$^{-1}$ similarly arises from optic (O) phonons. Inelastic neutron scattering measurements of the acoustic phonon dispersion in LaS[32] allow us to determine the origins of the acoustic Raman features. The 86 cm$^{-1}$ peak arises primarily from transverse (TA) and longitudinal (LA) acoustic phonons at the X and K critical points in the Brillouin zone, while the shoulder at 108 cm$^{-1}$ comes mainly from L point TA and LA phonons. The distinct O peak at 261 cm$^{-1}$ comprises contributions from both TO and LO phonons, with the highest frequency contributions from LO modes[2]. The broad and very weak bands occurring at frequencies higher than the optic phonon peak at 261 cm$^{-1}$ are second-order in origin reflecting primarily combinations of the disorder-induced first-order DOS Raman peaks. For example, the peaks at 347, 448 and 518 cm$^{-1}$ can be attributed to the combinations of the 86 and 108 (A) cm$^{-1}$ peaks with the 261 (O) cm$^{-1}$ peak (A+O), 2A/O+O and 2O, respectively (see Fig.22).

## B.  Growth of LaS nanoballs and nanoclusters

Nanoclusters of LaS were fabricated by pulsed laser ablation (PLA) of a LaS target (5 cm diameter) with the output of a KrF ($\lambda$=248 nm) excimer laser.  The laser pulse duration was 25 ns, and the laser energy was 700mJ per pulse.  The target was rotated during deposition in order to expose a fresh LaS surface to each successive ablating laser pulse. Ablation was carried out in a static Ar background pressure of 1 Torr. Nanoclusters were deposited on a Si substrate (for SEM characterization) and on a TEM grid, both of which were situated 5 cm from the target. A total of four laser shots were used to fabricate the TEM samples, while the samples used for SEM characterization were prepared by ablating the LaS target for five minutes at a repetition rate of 10 Hz.

Synthesis of nanoclusters by PLA typically results in a bimodal size distribution[33]. The smaller (1 nm < diameter < 10 nm) nanoclusters are formed by recombination in the gas phase of ablated atomic and molecular species. Larger nanoclusters are directly ejected from the target by a process denoted as splashing[34], which can be understood by considering the nature of the species ejected from the target as a function of depth into the target. Species originating close to the target surface are exposed to the full laser energy and are ejected primarily as atoms; however, at great distances into the target, the laser energy is attenuated to such an extent that the target is heated rather than ablated. This results in the ejection of molten nanodroplets of liquid target material. Although splashing can be an inconvenience for deposition of epitaxial thin films, we use it to our advantage in nanocluster synthesis.

Presented in Figures 23(top) and 23(bottom) are SEM micrographs of LaS nanoclusters formed by PLA. The nanocluster seen in Figure 23(top) is spherical and has a diameter on the order of 300 nm. The spherical nature of the nanocluster, as well as its fairly large diameter, implies that this species was directly ejected from the target by splashing. The spherical nature of the feature also suggests that the particle was in the liquid state during part of its journey from target to substrate. Although difficult to discern from this single SEM image, it is observed quite often that these hot liquid nanodroplets become flattened as they strike the substrate such that their height (above the substrate) is less than their diameter. Such characteristics are more easily seen by AFM analysis.

Shown in Figure 23(bottom) is an agglomeration of LaS nanoclusters that range in size from ~ 60 nm to 200 nm. It is interesting to note that the smaller nanoclusters tend to have spherical shapes and appear to have fused slightly with their adjacent nanoclusters (again consistent with their being ejected as liquid nanodroplets). It is also interesting to note that the larger nanoclusters appear to have more angular features possibly suggesting that they were ejected from the target with temperatures that were insufficient to cause melting.

Shown in Figure 24 are TEM micrographs of LaS nanoclusters. The nanocluster seen in the left image has a diameter of ~ 60 nm and has a spherical shape which is partially obscured by the support grid on the left side of the nanocluster. The spherical shape suggests that this object was liquid during its journey from target to substrate. Shown on the right of Figure 24 is a nanocluster that presumably flattened upon impact with the TEM grid. It is interesting to note that this nanocluster has some angular features, which suggests the possibility that the nanocluster had undergone partial crystallization upon impact.

# VI. SUMMARY AND CONCLUSIONS

Typically, low WF surfaces are generally highly chemically reactive. As a result, cold cathodes made of these materials are very unstable. For the last ten years, we have developed a new class of field emitters based on rare-earth monosulfide thin films with turn out to be very reliable and durable. In fact, rare-earth monosulfides are unusual in many respects. Not only do they possess high chemical stability (melting temperatures above 2000 $^{o}$C), but they also display metallic conduction. So far, LaS thin films have been successfully grown using PLD on Si and InP substrates at room temperature and on MgO substrates at high temperature in the presence of a $H_2S$ background gas.

LaS/MgO thin films were used successfully as cold cathode emitters with measured emitted current densities as high as 50 A/cm$^2$. For both LaS/Si and LaS/MgO thin films, the effective WF of the micron size thin films was determined from field emission measurements using the Scanning Anode Field Emission Microscopy (SAFEM) technique. From the Fowler-Nordheim plots of the field emission characteristics of the thin films, an effective WF around 1 eV was extracted. The physical reasons for these highly desirable low WF properties were explained using a patchwork field emission (FE) model of the emitting surface in which nanocrystals of low WF materials having a <100> orientation perpendicular to the surface and outcropping it are surrounded by a matrix of amorphous materials with higher WF[27].

Recent numerical calculations of the patch-field distribution across these nanopatchwork surfaces show that low work function nanosize zones (with diameter less than few tens of nm) are intrinsically protected by an electrostatic screen induced by the surrounding media with a higher WF. The latter prevents molecules from absorbing onto the lower WF areas protecting them from contamination until a bias applied to an anode in close proximity opens up channels for efficient field emission on top of the 1eV nanocrystals. Therefore, the nanopatch structure allows the emission of nondiffracted parallel nanosize electron (e-) beams with well defined spatial locations. The production of a high density array of these parallel e-beams is, therefore, no longer hindered by the presence of physical electrodes (like extracting lenses and diaphragms) in the very near neighborhood of a tip configuration cathode surface, as in current e-beam systems. With the patchwork cathode, the structure of the e-beam array and its resolution are only determined by the location and dimension of the nanopatches over a plane surface, opening a new route for coherent parallel e-beam applications.

More recently, we have been able to produce LaS nanoballs and nanoclusters by PLA. The latter can be used to control the density and size of the nanopatches to generate a high density of parallel nanosize electron-beams.

## ACKNOWLEDGMENTS

The work at the University of Cincinnati was supported by NSF grants ECS-990653, ECS-9632511, and collaborative GOALI-ECS-0523966. Part of this work was performed by M. Cahay under the Air Force Summer Fellowship Program, contract No. FA9550-07-C-0052, contract No. F49620-93-C-0063, and was also supported by Wright Laboratory. The work was also supported by the Air Force Research Laboratory, Sensors,

Directorate, of Wright-Patterson Air Force Base under contract No. F33615-98-C-1204. We thank J.-M. Picard for technical assistance with the Raman measurements.

**List of Tables**

Table 1. Materials Parameters of Some Sulfides of Rare-Earth Metals (cubic form): a (lattice constant in Å), WF (work function at room temperature), $T_m$ (melting point in °C), and $\rho$ electrical resistivity (in $\mu\Omega$cm )[5].

Table 2. Model used for fitting the ellipsometric data for a film of LaS on InP.

**Figure Captions**

Figure 1. Cross-section of the Friz cold cathode, which consists of an electron emitter (Au, Ag, or heavily doped InP substrate), a tens of nanometers thick wide-bandgap (2.4eV) CdS layer and a LaS thin film on the surface. The NEA at the CdS/LaS interface facilitates the injection of electron into vacuum. The top emitting surface is biased with an array of emitter fingers.

Figure 2. Energy band diagram across the Friz cathode from the substrate into vacuum. The substrate is held at ground potential and a positive bias is applied on the LaS thin film using the array of emitter fingers shown in Fig.1.

Figure 3. (Color online) (a) A pellet of the cubic phase of LaS about 1.5 cm long, 4 mm wide and 2 mm thick prepared by the carbon reduction process. (b) Optical micrograph showing a large golden platelet (about 30 $\mu m^2$) of the cubic phase of LaS with a staircase-like growth pattern.

Figure 4. Comparison of XRD scans of a LaS bulk sample (a) before and (b) after the carbon reduction process. In the latter, the Bragg reflection peaks corresponding to the oxysulfide ($La_2O_2S$) impurity phase have disappeared.

Figure 5. (Color online) Thin film of LaS deposited by PLD on a square inch Si substrate. The reflection of the finger shows the metallic character of the thin film.

Figure 6. (Color online) AFM scans over a 1x1 $\mu m^2$ area of a LaS thin film (about one micron thick) grown on a (001) Si substrate  The root mean square variation of the surface roughness is 1.74 nm over that area.

Figure 7. X-ray diffraction pattern at three different grazing angles of incidence ($\varphi$) from a LaS thin film deposited on a (100) Si substrate. Miller indices (hkl) of the reflections from the cubic rocksalt phase of LaS are identified. The peak located around $56^o$ is due to the (311) Bragg reflection observed for a bare Si substrate (Joint Committee on Powder Diffraction Standard card No. 77-211).

Figure 8. High-resolution TEM image of the LaS/Si interface for a 1 $\mu m$ thick LaS film grown on a (100) Si substrate. Despite the large lattice mistmatch between the two materials (8%), the interface appear to be rather smooth at the length scale shown in this TEM picture.

Figure 9. (Color online) AFM scans over a 2x2 $\mu m^2$ area of LaS thin-film grown on (a) (001) MgO substrate and (b) on a (100) Si substrate. The PLD parameters for thin-film growth were as follows: laser repetition rate of 4 Hz, deposition time of 30 minutes, and the substrate to target separation of 10 cm. In (a) the substrate temperature was 400°C with 24 mTorr of $H_2S$ background pressure while in (b) Si substrate was held at room temperature and no $H_2S$ background gas.

Figure 10. (Color online) AFM scan over a 2x2 $\mu m^2$ area of (a) a LaS thin film grown on a (001) MgO substrate. The PLD parameters were: substrate temperature of 500$^o$C, H$_2$S background pressure of 45 mTorr, laser repetition rate of 8 Hz, and a substrate to target separation of 10 cm. The top surface is composed of large platelets with size averaging near 293 nm.

Figure 11. UPS spectra of a LaS film deposited by pulsed laser ablation. The inset shows a blown up region at the Fermi energy (E$_F$).

Figure 12. (Color online) I-V characteristics for LaS thin-film and nanoprotruded cathodes. Here V$_{film}$ and V$_{nano}$ represent the actual applied potentials to obtain field emission. For the nanoprotruded dome the potential is found nearly 3.6 less than for a flat film to obtain the same field emission current. In other words, the field enhancement factor for the nanodome geometry is at least a factor of 3.6 higher.

Figure 13. (Color online) I-V characteristics for LaS deposited on MgO and on Si. The two sets of data were obtained at the same probe distances d, that means for the same field emission fields.

Figure 14. (Color online) (Top) HRTEM image of the deposited LaS film showing the flat surface and the nanocrystalline nature of the layer. (Bottom) Schematic representation of a patchwork FE through nanocrystallites of a LaS thin film in close proximity of the probe ball (anode with a spherical shape) used in SAFEM experiments. The nanocrystallites (a) have a low work function (crystallites (a) and (b) have the same orientations) and the nanocrystallites (b) are not field emitting because they are embedded in the layer; the lines surrounding the nanocrystallites (a) schematically represent the current lines collecting the emitted electrons.

Figure 15. (Color online) **(a)**: Schematic of a similar patch area with a low work function $\Phi_1$ and with a few nm diameter surrounded by a larger work function $\Phi_2$ amorphous phase. The separation between the two regions is not abrupt and the work function linearly increases from $\Phi_1$ to $\Phi_2$ over a connecting layer of width $L_C$ around 0.5 nm. **(b):** Schematic of potential energy profile at equilibrium at the interface between the central patch with low work function $\Phi_1$ and the surrounding amorphous material with larger work function $\Phi_2$.

Figure 16. (Color online) Evolution of the potential distribution over a 2 nm diameter patch showing the progressive opening over the patch for increasing applied field $F_{app}$. The patch has a WF of 1 eV and the surrounding medium a WF of 2.8 eV. The data in the figures represent the lowest surface barrier height of the opening. In figure (d), for example, for $F_{app}$ of 1 V/nm the surface barrier height at the center of the patch is 1.32 eV.

Figure 17. (Color online) Potential opening over the patch for two patch dimensions (1 and 4 nm) and for the same applied fields (0.1 and 0.5 V/m).

Figure18. Plots of I-V characteristics respectively for a patch cathode and for two uniform surface cathodes.

Figure 19. X-ray diffraction pattern at three different grazing angles of incidence from a LaS thin film deposited on a (100) InP substrate. Miller indices (hkl) of the reflections from the cubic rocksalt phase of LaS are identified.

Figure 20. (a,b) Cross-section TEM images of a LaS thin film on an InP substrate showing that there are two LaS layers on the InP substrate, and the thicknesses of LaS(1) and LaS(2) are 52 nm and 121 nm, respectively. Selected area diffraction pattern (c) from the LaS layers reveal that the LaS film has a poly-crystalline nature.

Figure 21. Ellipsometric spectra $\Psi(E)$ and $\Delta(E)$ measured from LaS on InP are shown in frames (a) and (b), respectively. The extracted complex permittivity dispersion curves $\varepsilon_1(E)$ and $\varepsilon_2(E)$ are shown in frame (c). Frame (d) shows the estimated penetration depth of the light into the LaS film based on the permittivity dispersion.

Figure 22. (Color online) The room temperature Raman spectrum of bulk LaS (from a target similar to that used in the growth chamber) compared with two LaS films grown on (100) InP. A and O refer to acoustic and optical phonons involved in first-order and second-order Raman scattering.

Figure 23. FE-SEM images of (a) single and (b) a cluster of LaS nanoballs grown on a Si substrate using the following PLA parameters: 248 nm excimer laser, laser pulse duration 25 ns, 700 mJ/pulse, 10 Hz, 5 min deposition, 1 Torr Ar background pressure, target to substrate separation 5 cm.

Figure 24. TEM micrographs of single LaS nanoparticle deposited on a TEM grid. The PLA parameters were: 248 nm excimer laser, laser pulse duration 25 ns, 700 mJ/pulse, 4 pulses, 1 Torr Ar background pressure, target to substrate separation 5 cm.

|  | ErS | YS | NdS | GdS | PrS | CeS | LaS | EuS | SmS |
|---|---|---|---|---|---|---|---|---|---|
| Lattice constant (Å) | 5.424 | 5.466 | 5.69 | 5.74 | 5.747 | 5.778 | 5.854 | 5.968 | 5.970 |
| Work function (300K) |  |  | 1.36 |  | 1.26 | 1.05 | 1.14 |  |  |
| Melting point ($^0$C) |  | 2060 | 2200 |  | 2230 | 2450 | 2200 |  | 1870 |
| Electrical resistivity ($\mu\Omega$) |  |  | 242 |  | 240 | 170 | 25 |  |  |

**Table 1**

| Top Layer | Rough oxidized layer | 35.4 nm |
|---|---|---|
| Layer 2 | Top LaS layer | 134.7nm |
| Layer 1 | Bottom LaS layer | 52.0 nm |
| Layer 0 | InP substrate | 0.5 mm |

**Table 2**

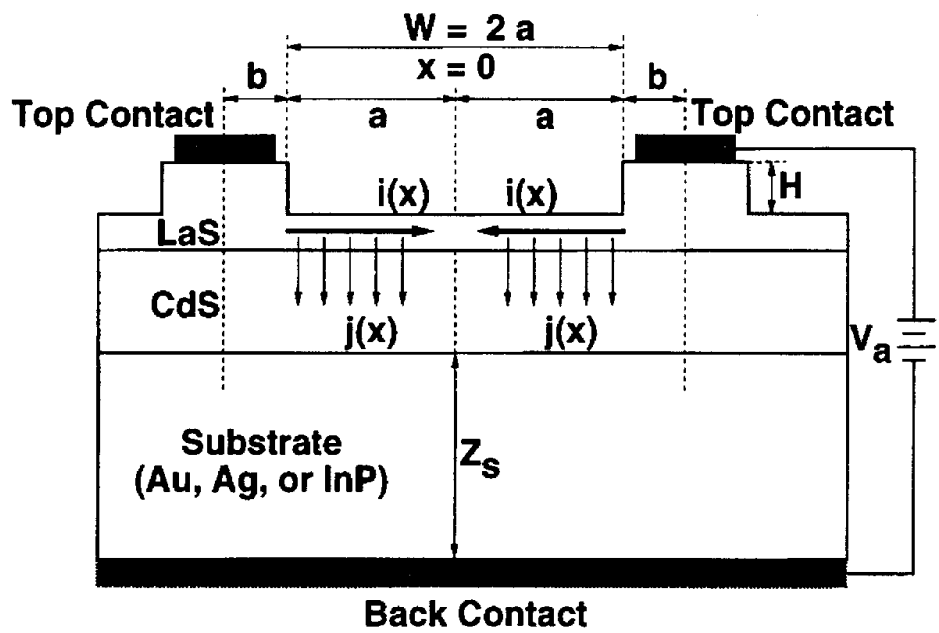

Figure 1 (M. Cahay et al.)

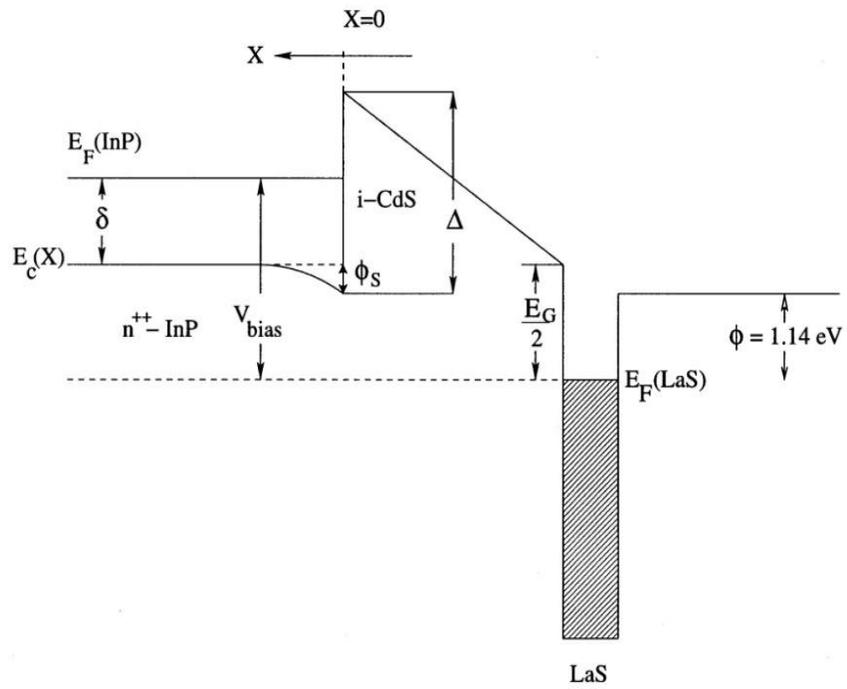

**Figure 2 (M. Cahay et al.)**

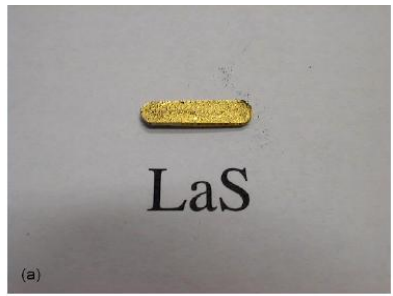

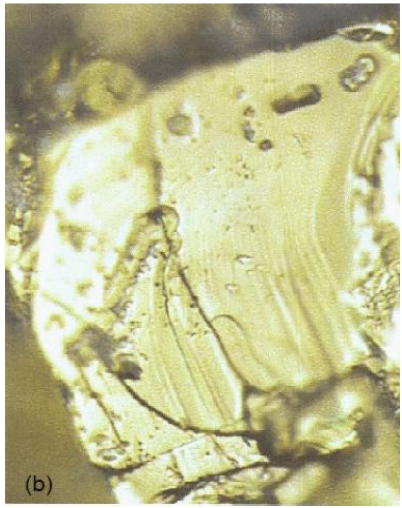

**Figure 3 (M. Cahay et al.)**

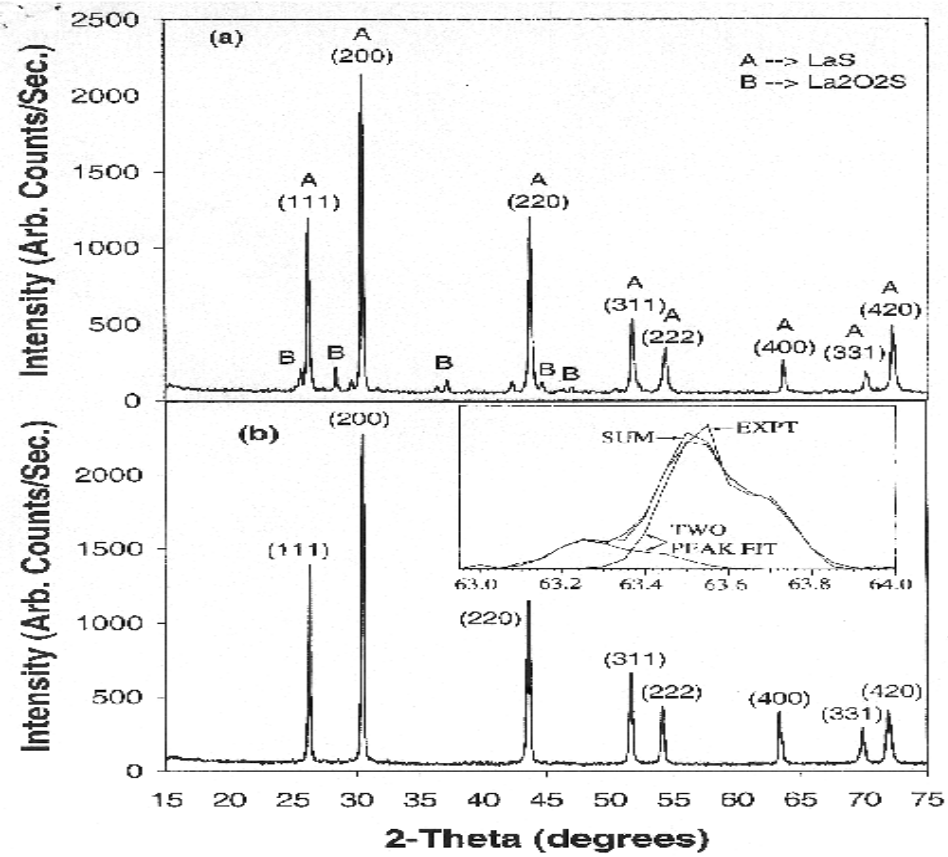

**Figure 4 (M. Cahay et al.)**

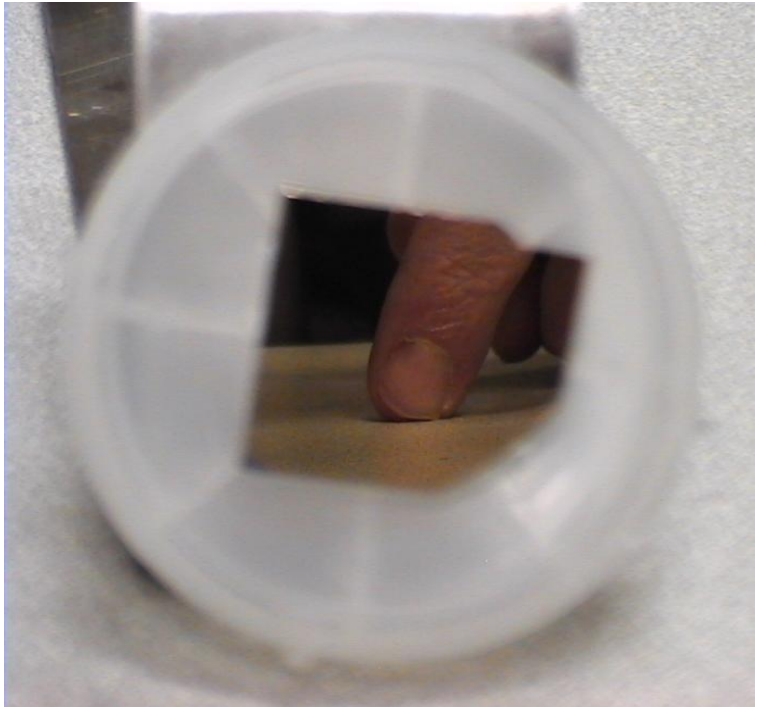

**Figure 5 (M. Cahay et al.)**

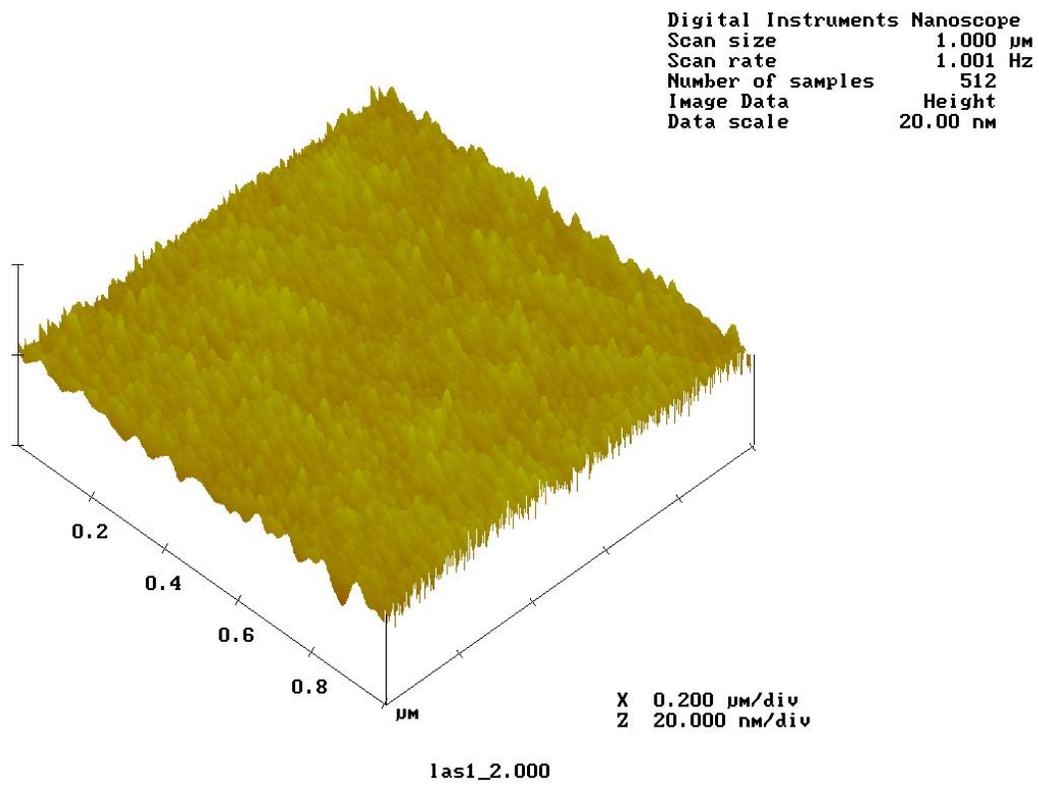

**Figure 6 (M. Cahay et al.)**

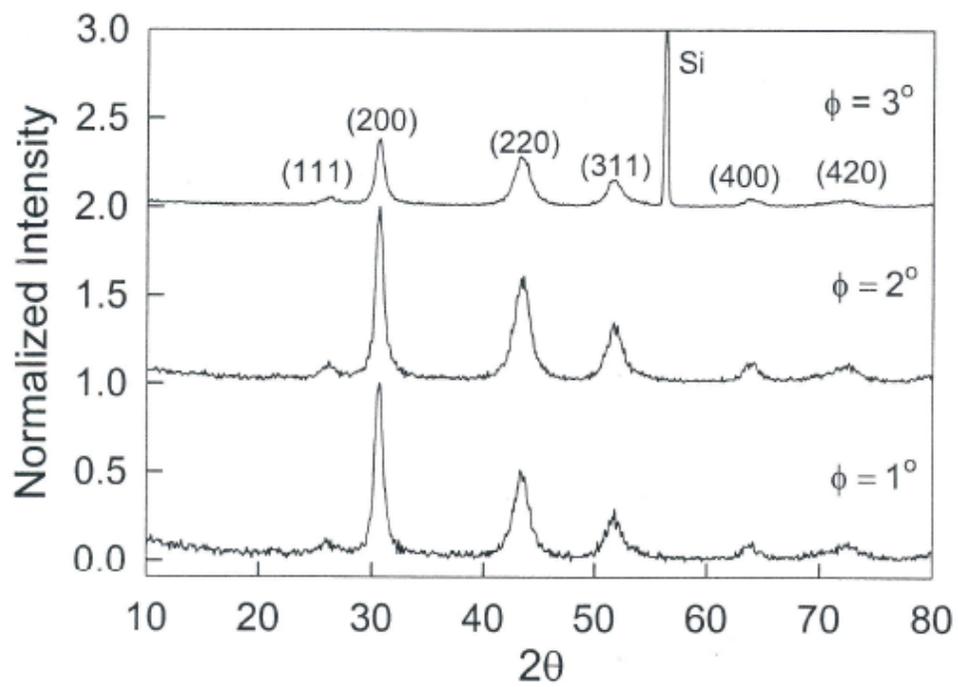

**Figure 7 (M. Cahay et al.)**

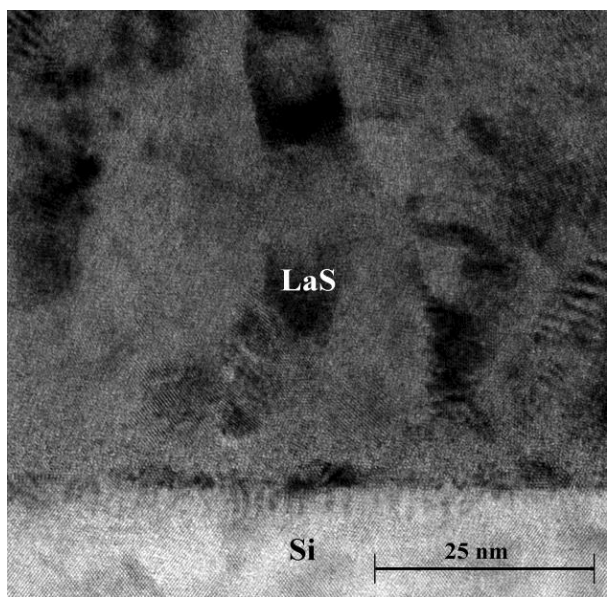

**Figure 8 (M. Cahay et al.)**

(a)

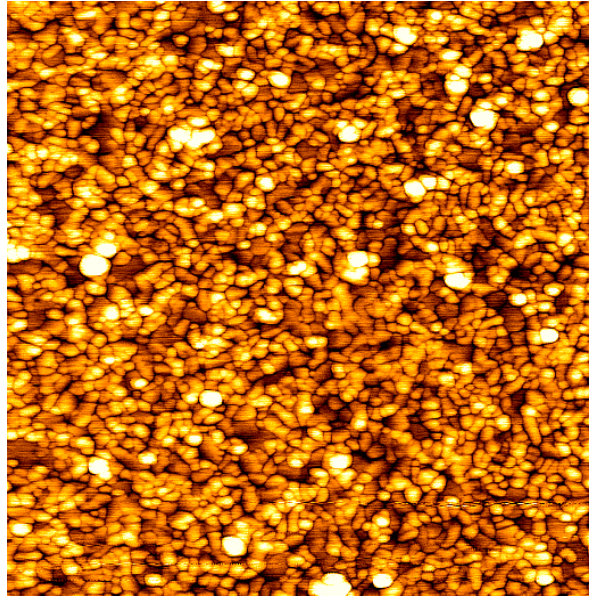

(b)

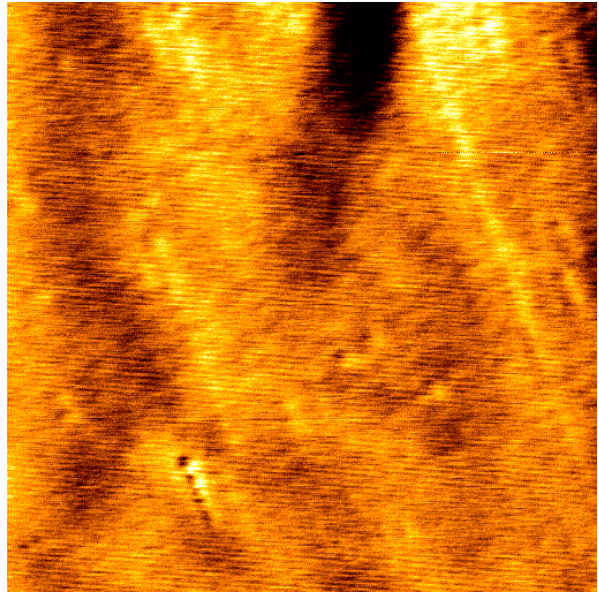

**Figure 9 (M. Cahay et al.)**

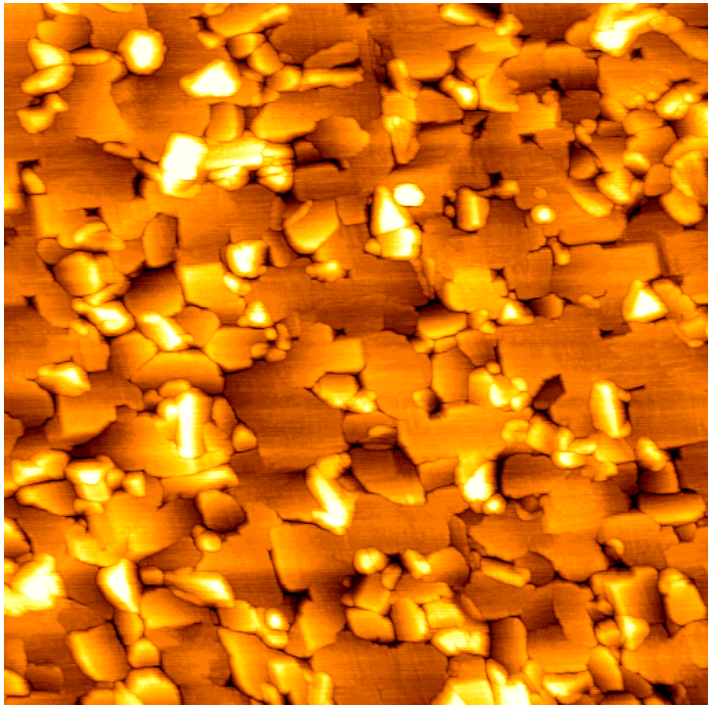

**Figure 10 (M. Cahay et al.)**

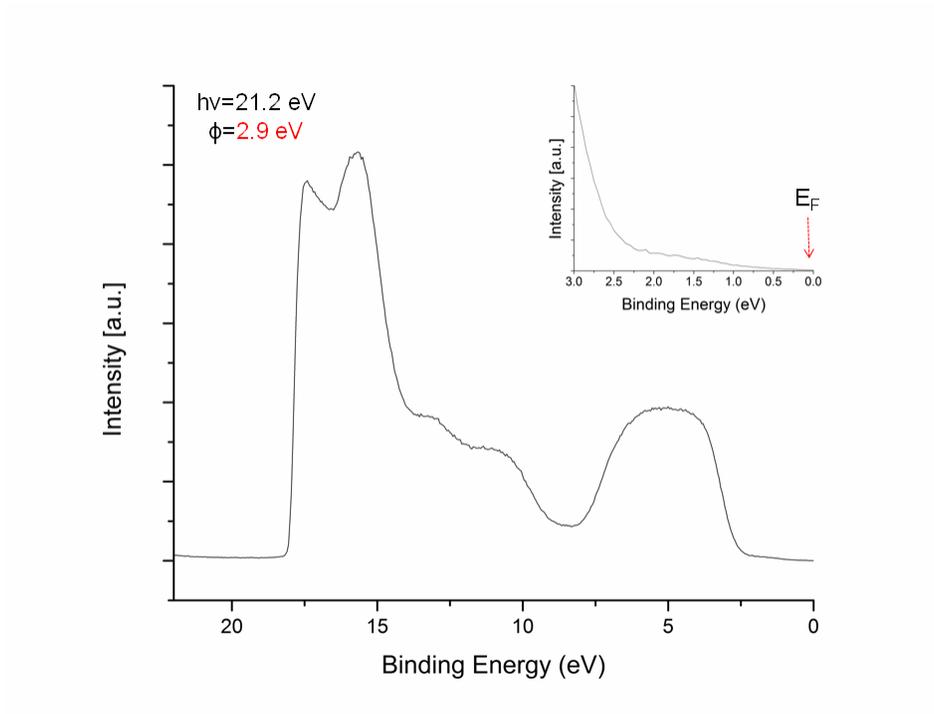

**Figure 11 (M. Cahay et al.)**

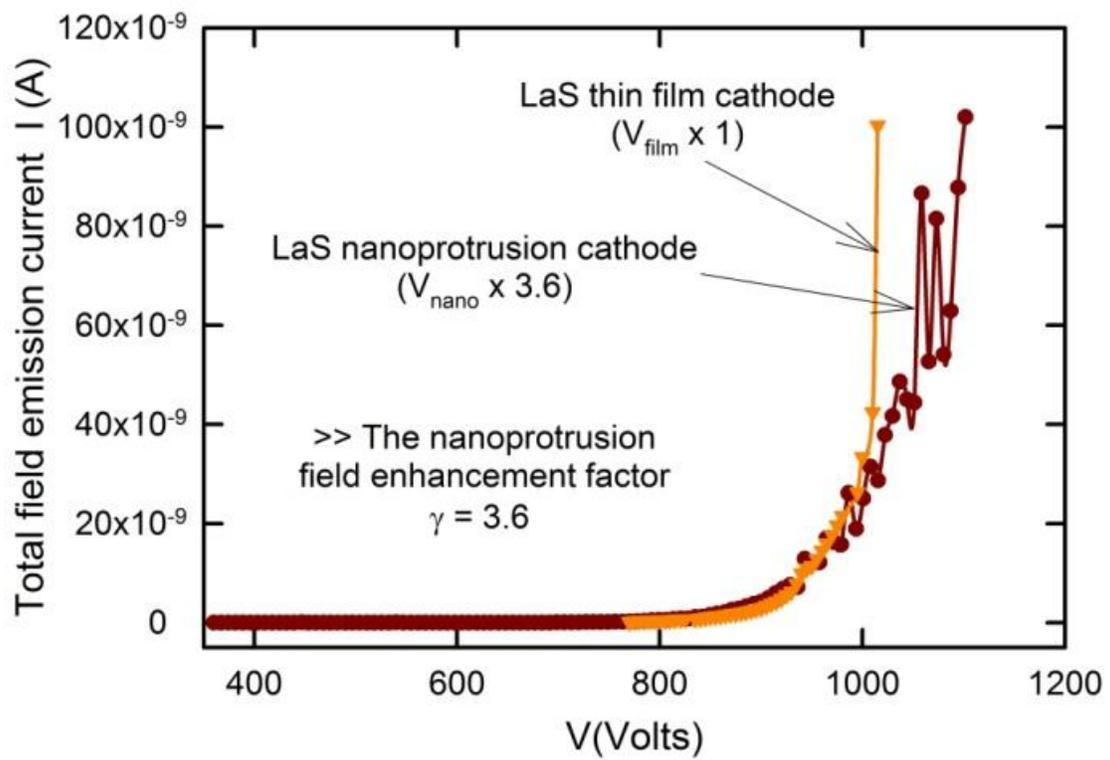

**Figure 12 (M. Cahay et al.)**

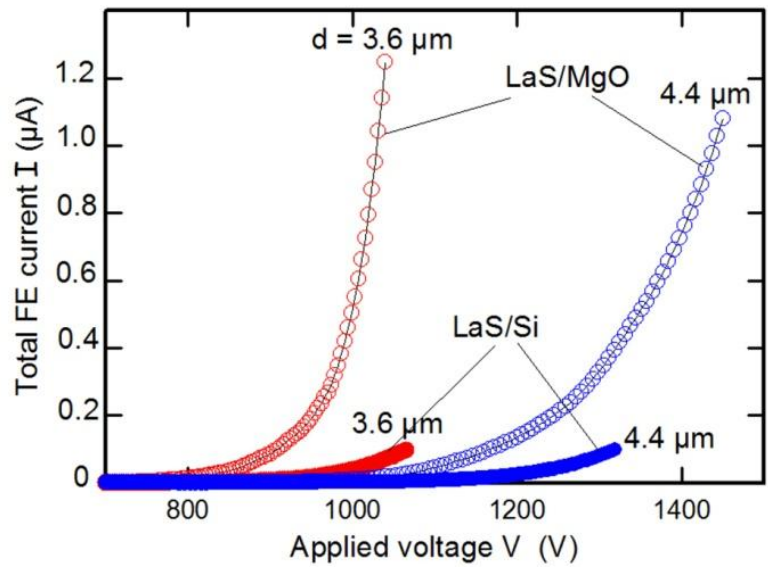

**Figure 13 (M. Cahay et al.)**

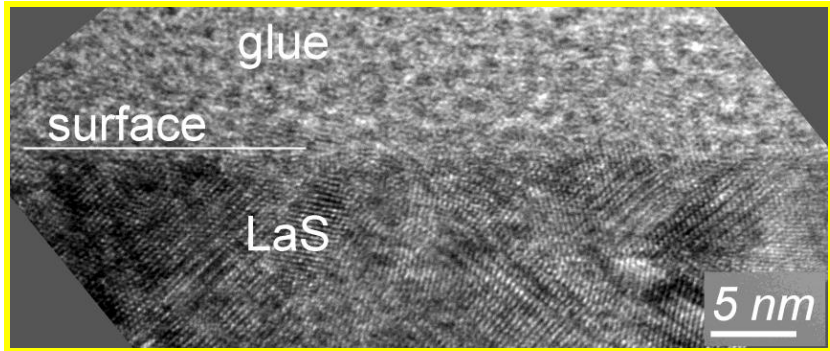

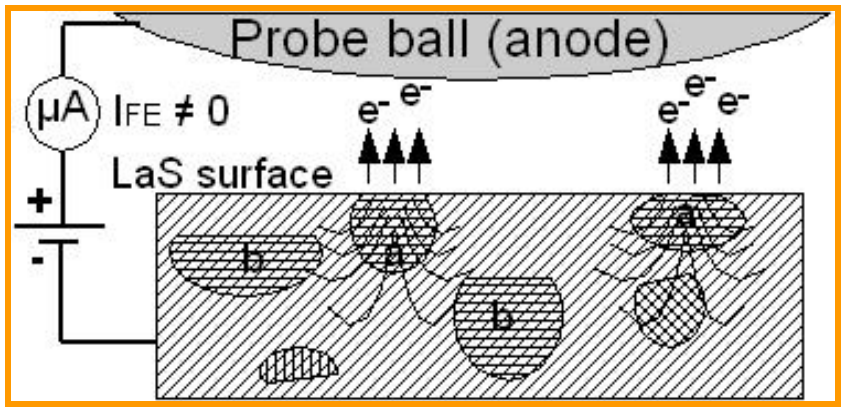

**Figure 14 (M. Cahay et al.)**

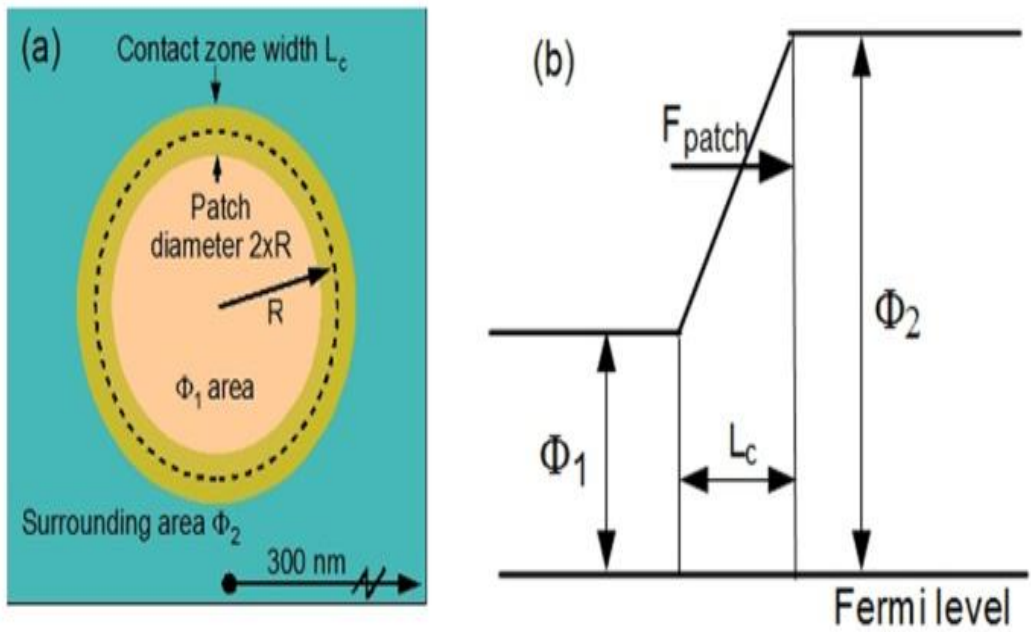

**Figure 15 (M. Cahay et al.)**

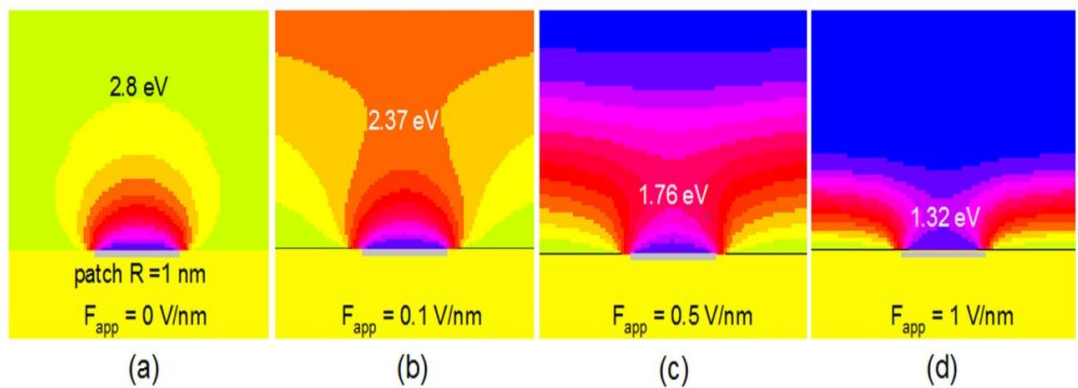

**Figure 16 (M. Cahay et al.)**

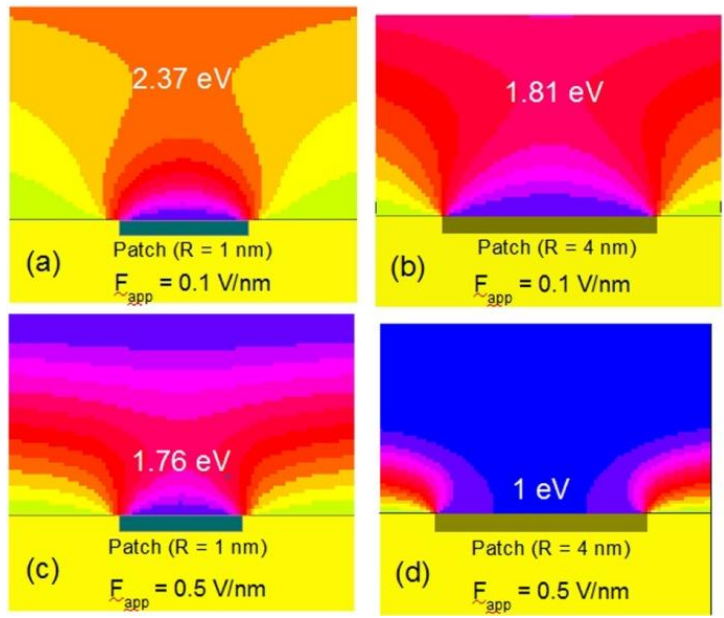

**Figure 17 (M. Cahay et al.)**

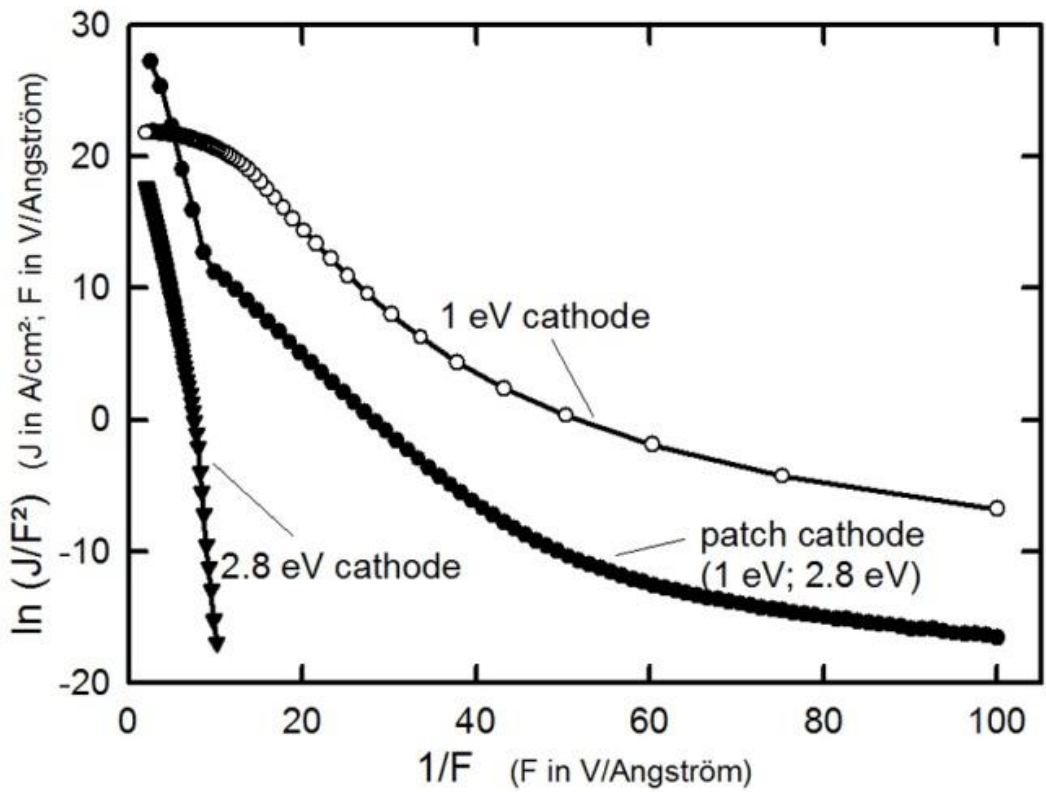

**Figure 18 (M. Cahay et al.)**

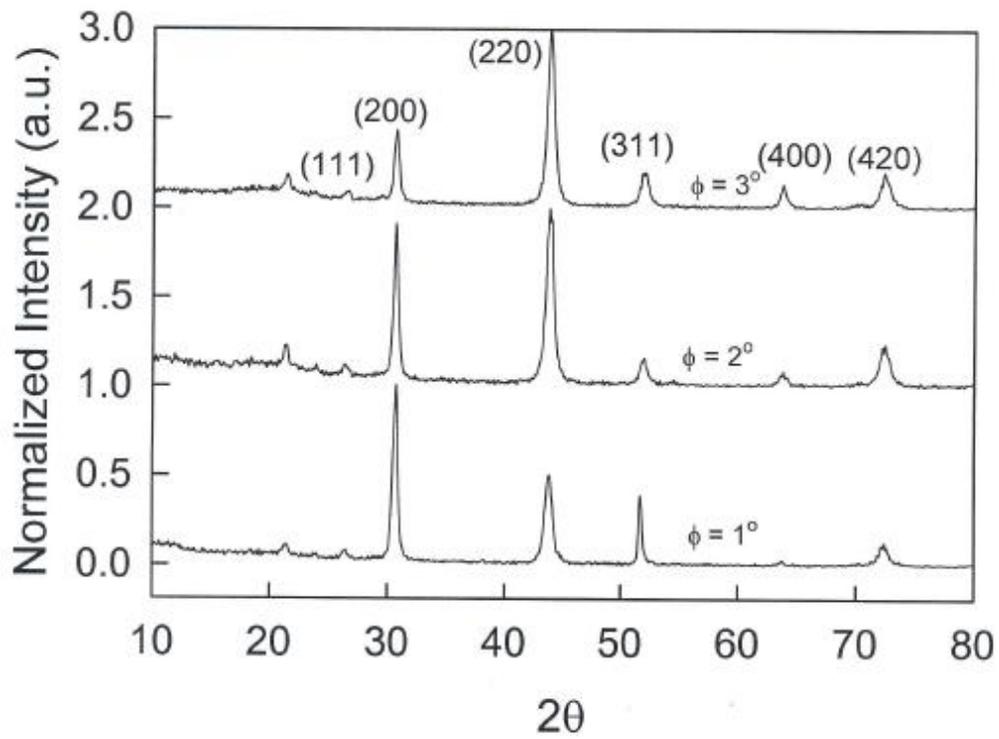

**Figure 19 (M. Cahay et al.)**

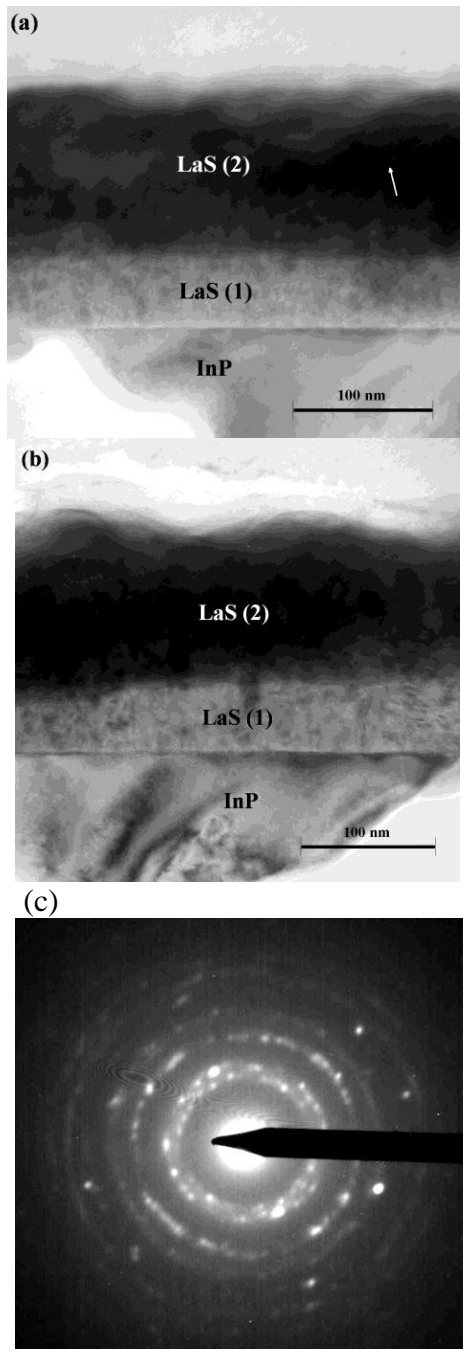

**Figure 20 (M. Cahay et al.)**

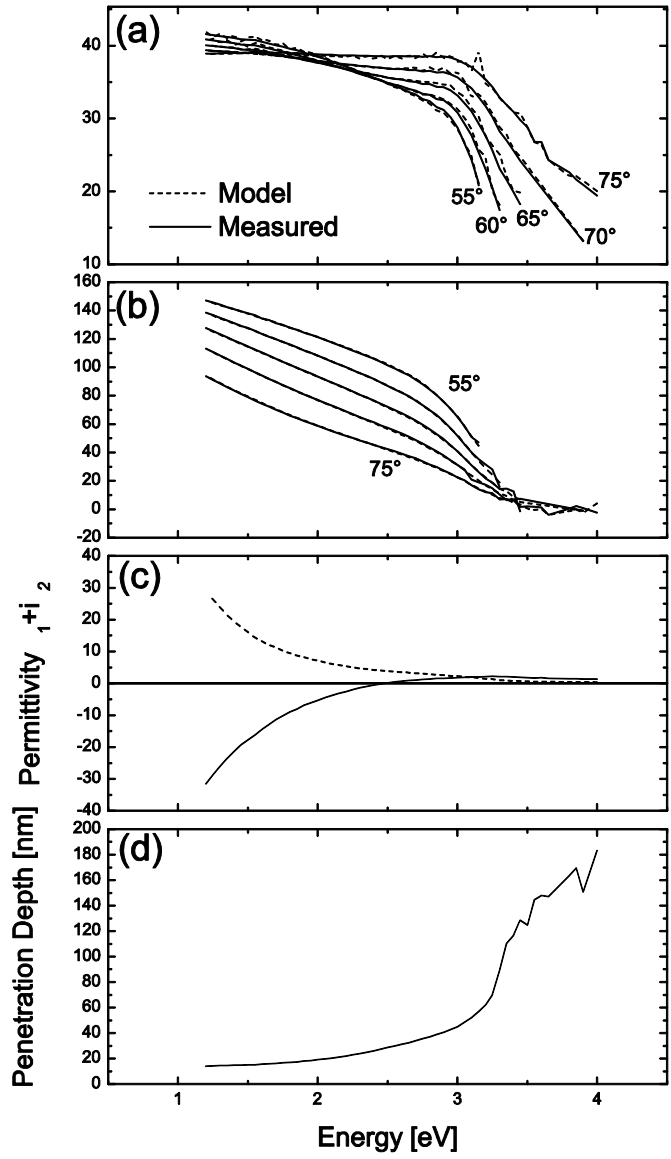

Figure 21 (M. Cahay et al.)

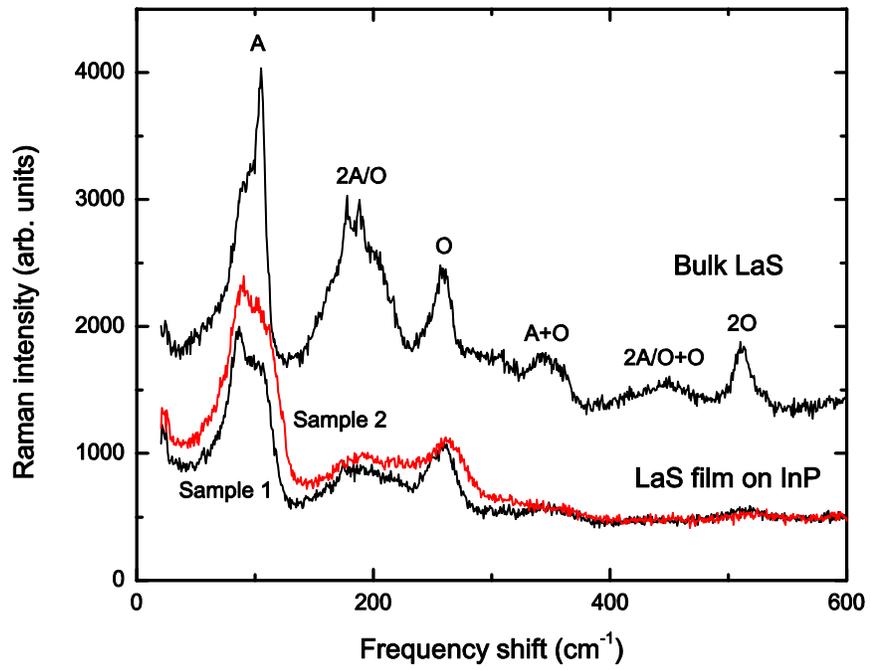

**Figure 22 (M. Cahay et al.)**

**(a)**

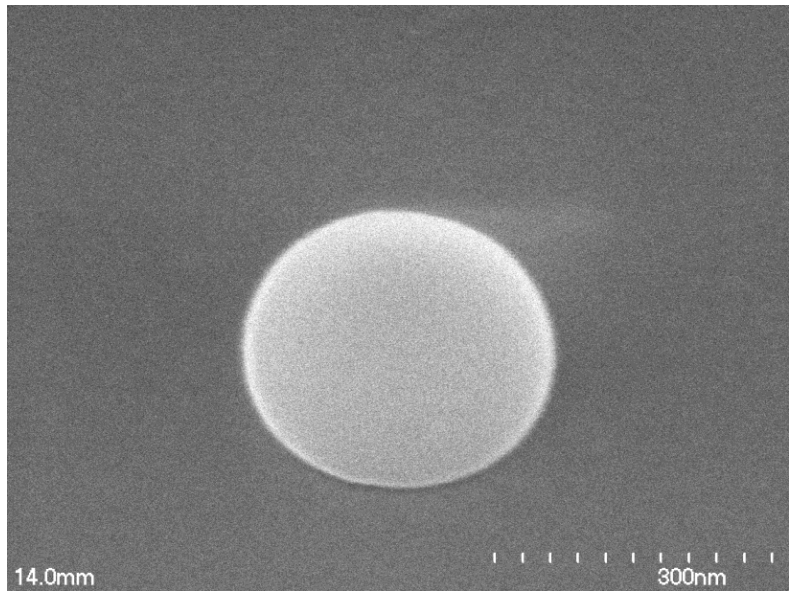

**(b)**

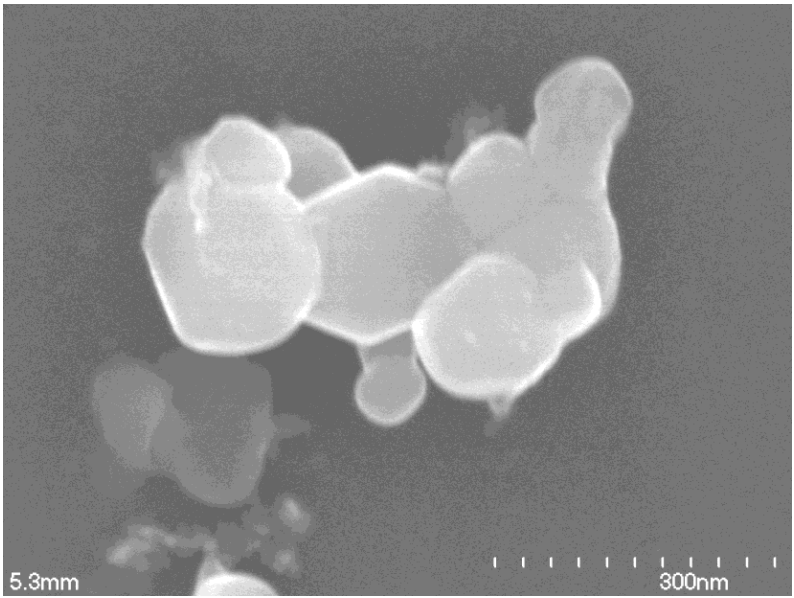

**Figure 23  (M. Cahay at al.)**

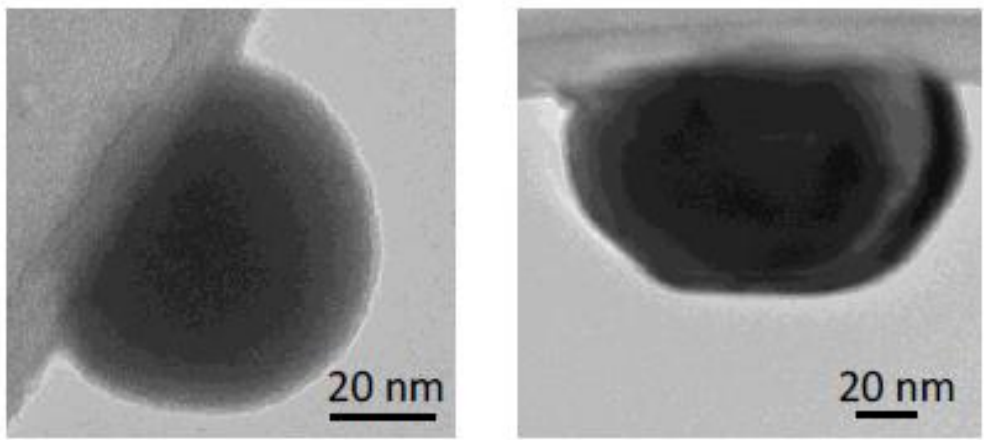

**Figure 24 (M. Cahay at al.)**